\newcommand{\be}{\begin{equation}}\newcommand{\ee}{\end{equation}}
\newcommand{\bea}{\begin{eqnarray}}\newcommand{\eea}{\end{eqnarray}}
\newcommand{\nn}{\nonumber\\[6pt]}
\newcommand{\p}[1]{(\ref{#1})}

\newcommand{\bD}{\overline D}
\newcommand{\cD}{{\cal D}}

\newcommand{\bT}{{\overline T}}
\newcommand{\bV}{{\overline V}}
\newcommand{\bQ}{{\overline Q}}
\newcommand{\bS}{{\overline S}}
\newcommand{\bt}{{\bar\theta}}
\newcommand{\bxi}{{\bar\xi}}
\newcommand{\bpsi}{{\bar\psi}}

\newcommand{\bphi}{{\bar\phi}}
\newcommand{\blam}{{\bar\lambda}}
\newcommand{\bLam}{{\overline\Lambda}}
\newcommand{\bomega}{{\bar\omega}}
\newcommand{\beps}{{\bar\epsilon}}

\newcommand{\tV}{{\tilde V}}

\newcommand{\bX}{{\overline X}}
\newcommand{\bZ}{{\overline Z}}
\newcommand{\sfrac}[2]{{\textstyle\frac{#1}{#2}}}

\documentclass[12pt]{article}
\usepackage{amscd,amsmath,amssymb}

\begin{document}

\thispagestyle{empty}
\vspace{2cm}
\begin{flushright}
hep-th/0310299 \\
ITP-UH-09/03 \\[25mm]
\end{flushright}
\begin{center}
{\Large\bf N=4, d=1 supermultiplets from} \\
\vspace{0.3cm}

{\Large\bf nonlinear realizations of D(2,1;$\alpha$)}
\end{center}
\vspace{1cm}

\begin{center}
{\large\bf  E. Ivanov${}^{a}$,
S. Krivonos${}^{a}$, O. Lechtenfeld${}^{b}$ }
\end{center}

\begin{center}
${}^a$ {\it Bogoliubov  Laboratory of Theoretical Physics, JINR, 141980 Dubna,
Russia}

\vspace{0.2cm}
${}^b$ {\it Institut f\"ur Theoretische Physik, Universit\"at Hannover,} \\
{\it Appelstra\ss{}e 2, 30167 Hannover, Germany}
\end{center}

\begin{center}
{\tt eivanov, krivonos@thsun1.jinr.ru, lechtenf@itp.uni-hannover.de}
\end{center}
\vspace{2cm}

\begin{abstract}
\noindent
Proceeding from nonlinear realizations of the most general $N{=}4$, $d{=}1$
superconformal symmetry associated with the supergroup $D(2,1;\alpha)$,
we construct all known and two new  off-shell $N{=}4$, $d{=}1$ supermultiplets
as properly constrained $N{=}4$ superfields. We find plenty of nonlinear
interrelations between the multiplets constructed and present a few examples
of invariant superfield actions for them. The superconformal transformation
properties of these multiplets are explicit within our method.

\end{abstract}

\newpage
\setcounter{page}{1}
\section{Introduction}
Supersymmetric quantum mechanics (SQM), being the simplest supersymmetric theory,
has a lot of
interesting and important applications. In particular, it provides a nice
laboratory for
the study of many characteristic features of supersymmetric theories in higher
dimensions
(see e.g. \cite{rev1} and refs. therein). Superconformally invariant SQM models
are of special
significance in the AdS$_2/$CFT$_1$ correspondence and black hole moduli spaces
\cite{rev2}.

The interest in SQM models with extended supersymmetry is
largely due to the fact that extended
supersymmetry in one dimension displays a number of surprising peculiarities
as compared with
its higher-dimensional counterparts. For instance, $N{\geq} 4$ supersymmetric
models feature
target space geometries which in general cannot be reproduced from those
of higher-dimensional
models via dimensional reduction \cite{rev3}. Another phenomenon
(tightly related to
the just mentioned one) is the diversity of off-shell multiplets
of $N{\geq} 4$ supersymmetries in $d{=}1$.
Some of them can be directly recovered by dimensional
reduction from off-shell multiplets of, say, $N{=}1$ or $N{=}2$ supersymmetries
in $d{=}4$, but
there exist some off-shell multiplets whose $d{=}4$ analogs
are on shell! A striking
example is the $N{=}4, d{=}1$ multiplet $({\bf 4, 4, 0})$ \cite{{CP},{GPS}}
that comprises four bosonic
and four fermionic physical degrees of freedom but contains no auxiliary fields
at all. Its analog
in $d{>}1$ is the on-shell hypermultiplet.

In view of this great number of inequivalent off-shell multiplets
for $N{\geq} 4$ supersymmetries in
$d{=}1$ it is desirable to work out a self-consistent method of deriving
the appropriate
superfields and their
irreducibility constraints directly in one dimension, without resorting to
the dimensional reduction
procedure. To have such a tool and the complete list of off-shell superfields
obtained with its help
is important both for constructing new SQM models and for establishing precise
interrelations between
them.

In the present paper we focus on the case of $N{=}4, d{=}1$ supersymmetry
(with 4 real
supercharges) and propose to derive its various irreducible off-shell
superfields from different
nonlinear realizations of the most general $N{=}4, d{=}1$ superconformal
group $D(2,1;\alpha)$ \cite{FRS}.
An advantage of this approach is that it simultaneously specifies the
superconformal transformation
properties of the superfields, though the latter can equally be used for
constructing non-conformal
SQM models as well. As the essence of these techniques, any given
irreducible $N{=}4, d{=}1$ superfield
comes out
as a Goldstone superfield parametrizing, together with the
$N{=}4, d{=}1$ superspace coordinates,
some supercoset of $D(2,1;\alpha)$. The method was
already employed in our paper \cite{IKL}
where we have re-derived the off-shell multiplet $({\bf 3, 4, 1})$
\cite{{ismi},{bepa},{stro2}} from
the nonlinear realization
of  $D(2,1;\alpha)$ in the coset with an
$SL(2,R)\times \left[SU(2)/U(1)\right]$ bosonic part
(the second $SU(2)\subset D(2,1; \alpha)$ was placed into the stability subgroup).

Here we consider nonlinear realizations of the same conformal
supergroup  $D(2,1;\alpha)$
in its other coset superspaces. In this way we reproduce
the $({\bf 4, 4, 0})$ multiplet
and also derive two new off-shell multiplets which, to the best of
our knowledge, have not been used
before in $N{=}4, d{=}1$ SQM model building.
The $({\bf 4,4,0})$ multiplet is represented by superfields
parametrizing a supercoset with the bosonic part being $SL(2,R)\times SU(2)$,
where the dilaton and the three parameters
of $SU(2)$ are identified with the four physical bosonic fields.
One of the new Goldstone multiplets is a $d{=}1$ analog of the so-called
nonlinear multiplet of $N{=}2, d{=}4$ supersymmetry \cite{nlin}.
It has the same off-shell contents $({\bf 3,4, 1})$ as the multiplet
employed in \cite{IKL}
but it obeys a different constraint and enjoys different superconformal
transformation properties. It corresponds to the specific nonlinear
realization of
$D(2,1; \alpha)$ where the dilatation generator and one of the two $SU(2)$
subgroups are placed
into the stability subgroup. One more new multiplet of similar type
is obtained by placing into the stability subgroup,
along with the dilatation and three $SU(2)$ generators,
also the $U(1)$
generator from the second $SU(2) \subset D(2,1;\alpha)$. It has the same
field content as
a chiral $N{=}4, d{=}1$ multiplet, i.e. $({\bf 2, 4, 2})$. Hence, it may be termed
the nonlinear chiral
supermultiplet. It is exceptional in the sense that no analogs of it are known
in $N{=}2, d{=}4$
superspace. For the two new multiplets, we construct general off-shell
superfield actions and consider
in some detail (by passing to components) a few instructive
examples for these
actions. Also we find a number of surprising interrelations between the
superfields
in question. Finally, we apply our techniques to a central
charge-extended
$SU(1,1\vert 2)$ supergroup obtained as a contraction of $D(2,1;\alpha)$ and
derive the appropriate analogs of $N{=}4, d{=}1$ multiplets considered before.

\section{Supergroup D(2,1;$\alpha$) and its  nonlinear realizations}
We use the standard definition  of the superalgebra $D(2,1;\alpha)$ \cite{FRS}
with the notations of ref. \cite{IKL}.
It contains nine bosonic generators which form a direct sum of $sl(2)$
with generators
$P,D,K$ and two $su(2)$ subalgebras with generators
$V, \bV, V_3\, \; \mbox{ and }
\; T, \bT, T_3$, respectively:
\bea\label{alg1}
&& i\left[ D,P\right] =P,\;  i\left[ D,K\right]=-K ,\;
i\left[ P,K\right]=-2D , \quad
i\left[ V_3,V\right]=-V,\;  i\left[ V_3,\bV \right]=\bV, \nn
&& i\left[ V,\bV\right]=2V_3,\quad
i\left[ T_3,T\right]=-T,\;  i\left[ T_3,\bT \right]=\bT,\;
i\left[ T,\bT\right]=2T_3.
\eea
The eight fermionic generators $Q^i,\bQ_i,S^i,\bS_i$
are in the fundamental representations of all bosonic subalgebras
(in our notation only
one $su(2)$ is manifest, viz. the one with generators
$V, \bV, V_3$):
\bea\label{alg2}
&&i\left[D ,Q^i \right] = \frac{1}{2}Q^i,\;
i\left[D ,S^i \right] = -\frac{1}{2}S^i, \quad
i\left[P ,S^i \right] =-Q^i,\;
i\left[K ,Q^i \right] =S^i, \nonumber \\
&& i\left[V_3 ,Q^1 \right] =\frac{1}{2}Q^1,\; i\left[V_3 ,Q^2 \right]
=-\frac{1}{2}Q^2,\quad
i\left[V ,Q^1 \right] =Q^2, \; i\left[V ,\bQ_2 \right] =-\bQ_1, \nonumber \\
&&i\left[V_3 ,S^1 \right] =\frac{1}{2}S^1,\;    i\left[V_3 ,S^2 \right]
=-\frac{1}{2}S^2, \quad
i\left[V ,S^1 \right] =S^2, \;    i\left[V ,\bS_2 \right] =-\bS_1,  \nonumber\\
&& i\left[T_3 ,Q^i\right] =\frac{1}{2}Q^i, \;
i\left[T_3 ,S^i\right] =\frac{1}{2}S^i, \quad
i\left[T ,Q^i\right] =\bQ^i, \;
i\left[T ,S^i\right] =\bS^i
\eea
(and c.c.). The splitting of the fermionic
generators into the $Q$ and $S$ sets is natural and useful, because
$Q^i,\bQ_k$ together with
$P$ form $N=4, d=1$ super Poincar\'e subalgebra, while $S^i,\bS_k $ generate
superconformal translations:
\be\label{allg3}
\left\{Q^i ,\bQ_j \right\} = -2\delta^i_j P , \quad \left\{S^i ,\bS_j \right\}
=-2\delta^i_j K .
\ee
The non-trivial dependence of the superalgebra $D(2,1;\alpha)$ on the parameter
$\alpha$
manifests itself only in the cross-anticommutators of the Poincar\'e and conformal
supercharges:
\bea\label{alg4}
&& \left\{ Q^i,S^j \right\} =-2(1+\alpha )\epsilon^{ij} \bT , \;
\left\{Q^1 ,\bS_2 \right\} =2\alpha \bV ,\;\left\{Q^1 ,\bS_1 \right\}
=-2D-2\alpha V_3+2(1+\alpha)T_3 ,
               \nonumber \\
&& \left\{Q^2 ,\bS_1 \right\} =-2\alpha V, \;\left\{Q^2 ,\bS_2 \right\}
=-2D +2\alpha V_3+2(1+\alpha)T_3
\eea
(and c.c.). The generators $P,D,K$ are chosen hermitian and the remaining ones obey
the following
conjugation rules:
\be\label{conjug}
\left( T \right)^\dagger = \bT, \; \left( T_3\right)^\dagger =-T_3 , \;
\left( V \right)^\dagger = \bV, \; \left( V_3\right)^\dagger =-V_3 , \;
\overline{\left( Q^i \right)}=\bQ_i,\; \overline{\left( S^i \right)}=\bS_i.
\ee

The parameter $\alpha $ is an arbitrary real number. At $\alpha = 0$ and
$\alpha = -1$ one of the
$su(2)$ algebras decouples and the superalgebra $su(1,1\vert 2)\oplus su(2)$
is recovered.
The superalgebra $D(2,1;1)$ is isomorphic
to $osp(4^*\vert 2)\,$.\footnote{ Sometimes $D(2,1; \alpha)$ is defined
so that these special values of $\alpha$ are excluded \cite{FRS}.
In our definition we retain these values in order to be able to
consider all inequivalent cases on equal footing.}
There are some equivalent choices of the parameter $\alpha$ which lead
to isomorphic
algebras $D(2,1;\alpha)$ \cite{FRS}.

We will be interested in diverse nonlinear realizations  of the
superconformal group $D(2,1;\alpha)$
in its coset superspaces. As the starting point we shall consider the following
parametrization of the supercoset
\be\label{coset}
g=e^{itP}e^{\theta_i Q^i+\bt^i \bQ_i}e^{\psi_i S^i+\bpsi^i \bS_i}
e^{izK}e^{iuD}e^{i\varphi V+ i\bar\varphi \bV}e^{\phi V_3}~.
\ee
The coordinates $t, \theta_i, \bt^i$ parametrize the $N=4, d=1$ superspace. All other
supercoset parameters are Goldstone $N=4$ superfields. The  group
$SU(2)\propto \left( V,\bV,V_3\right)$
linearly acts on the doublet indices $i$ of spinor coordinates and Goldstone
fermionic superfields,
while the bosonic Goldstone superfields $\varphi, \bar\varphi, \phi$ parametrize
this $SU(2)$.
Another $SU(2)$ as a whole is placed in the stability subgroup and acts only on
fermionic Goldstone superfields and $\theta$'s, mixing them with their conjugates.
With our choice of the $SU(2)$ coset we are led to assume that $\alpha \neq 0$.
We could
equivalently choose another $SU(2)$ to be nonlinearly realized and the first one
to belong
to the stability subgroup, then the restriction $\alpha \neq -1$ would be imposed.

In principle, we could lift up the second $SU(2)$ into the coset, with adding
the relevant
Goldstone superfields. In Sec. 8 we shall briefly discuss this more general situation
and
argue that it does not add new non-trivial options.

The left-covariant Cartan one-form $\Omega$ with values in the superalgebra
$D(2,1;\alpha)$
is defined by the standard relation
\be \label{cformdef}
g^{-1}\,d\,g = \Omega~.
\ee
In what follows we shall need the explicit structure of several
important one-forms in the expansion of $\Omega$ over the generators,
\bea\label{cforms}
&& \omega_D = idu-2\left( \bpsi^i d\theta_i + \psi_i d\bt^i \right)
  -2iz d{\tilde t} \; , \nn
&& \omega_V=\frac{e^{-i\phi}}{ 1 +\Lambda\bLam}\left[ id\Lambda+{\hat \omega}_V+
 \Lambda^2{\hat\bomega}_{V}-\Lambda{\hat\omega}_{V_3}\right], \;
 \bomega_{V}=\frac{e^{i\phi}}{ 1{+}\Lambda\bLam}\left[ id\bLam+{\hat \bomega}_{V}+
 \bLam^2{\hat\omega}_{V}+\bLam{\hat\omega}_{V_3}\right],\nn
&& \omega_{V_3}=d\phi+\frac{1}{ 1{+}\Lambda\bLam}\left[ i\left( d\Lambda\bLam -
\Lambda d\bLam\right)
+\left( 1{-}\Lambda\bLam\right){\hat\omega}_{V_3}
   -2\left( \Lambda{\hat\bomega}_{V}-\bLam{\hat\omega}_V \right)\right].
\eea
Here
\bea\label{cforms1}
&& {\hat\omega}_V=2\alpha \left[ \psi_2 d\bt^1 -\bpsi^1\left( d\theta_2-
       \psi_2 d{\tilde t}\right)\right], \;
 {\hat\bomega}_{V}=2\alpha \left[ \bpsi^2 d\theta_1 -\psi_1
    \left( d\bt^2-\bpsi^2 d{\tilde t}\right)\right],\nn
&& {\hat\omega}_{V_3}=2\alpha\left[ \psi_1 d\bt^1 -\bpsi^1 d \theta_1-
   \psi_2 d\bt^2 +\bpsi^2 d \theta_2 +
 \left( \bpsi^1 \psi_1 -\bpsi^2 \psi_2\right)d{\tilde t}\right],
\eea
\be
d{\tilde t} \equiv dt +i\left( \theta_id\bt^i+\bt^id\theta_i\right) \;,
\ee
and
\be \label{Lambda}
\Lambda = \frac{ \tan \sqrt{\varphi\bar\varphi}}{\sqrt{\varphi\bar\varphi}}\varphi
\; ,\;
\bLam = \frac{ \tan \sqrt{\varphi\bar\varphi}}{\sqrt{\varphi\bar\varphi}}\bar\varphi
\; .
\ee
The semi-covariant (fully covariant only under Poincar\'e supersymmetry)
spinor derivatives are defined by
\be
D^i=\frac{\partial}{\partial\theta_i}+i\bt^i \partial_t\; , \;
\bD_i=\frac{\partial}{\partial\bt^i}+i\theta_i \partial_t\; , \;
\left\{ D^i, \bD_j\right\}= 2i \delta^i_j \partial_t \; .
\ee

Let us quote the transformation properties of the $N=4$ superspace coordinates
and the basic Goldstone superfields under the transformations of supergroup
$D(2,1;\alpha)$.

The variations of the $N=4, d=1$ superspace coordinates
under the $N=4, d=1$ Poincar\'e supergroup
ar generated by acting on the coset element \p{coset} from the left by the element
\footnote{When summing over doublet indices we assume them to stay in a natural
position;
the Grassmann coordinates and their conjugates carry lower case and upper
case indices, respectively.
We use a short-hand notation
$\bar\psi \cdot \xi = \bar\psi^i \xi_i = - \xi\cdot \bar\psi$.
The contraction of spinors of equal kind is defined as $a\cdot b = a^ib_i$,
$\bar a\cdot \bar b = \bar a_i{\bar b}{}^i$.}
\be
g_0= e^{\varepsilon_i Q^i +\bar\varepsilon^i \bQ_i} \in D(2,1;\alpha)~.
\ee
The resulting transformations are
\be
\delta t = i\left(\theta\cdot\bar\varepsilon -\varepsilon\cdot\bar\theta\right)~, \quad
\delta \theta_i = \varepsilon_i~, \;\;\delta\bar\theta^i = \bar\varepsilon^i~.
\ee
All our superfields are scalars under the latter transformations.

The superconformal transformations are generated by acting on the coset
element \p{coset} from
the left by the element
\be\label{superconf}
g_1= e^{\epsilon_i S^i +\beps^i \bS_i}~,
\ee
and explicitly they read
\bea\label{superconf1}
&& \delta t=-it \left( \epsilon \cdot\bt +\beps\cdot\theta \right) +(1+2\alpha)
  \theta\cdot\bt \left( \epsilon\cdot\bt -\beps\cdot\theta \right),\nn
&& \delta \theta_i= \epsilon_i t -2i \alpha \theta_i (\theta\cdot \beps) + 2i
  (1+\alpha) \theta_i (\bt\cdot\epsilon) -i (1+2\alpha)
\epsilon_i (\theta\cdot\bt) \;, \nn
&& \delta u= -2i \left( \epsilon \cdot\bt +\beps\cdot\theta\right),\; \nn
&&\delta\phi=2\alpha\left[\beps^1\theta_1-\beps^2\theta_2 -
\epsilon_1\bt^1+\epsilon_2\bt^2
+\left(\beps^2\theta_1-\epsilon_1\bt^2\right)\Lambda+
\left(\beps^1\theta_2-\epsilon_2\bt^1\right)\bLam\right], \nn
&& \delta \Lambda= 2i\alpha\left[\theta_2\beps^1-\bt^1\epsilon_2+
\left(\bt^2\epsilon_1-\theta_1\beps^2\right)\Lambda^2+\left(\bt^1\epsilon_1-
  \theta_1\beps^1+\theta_2\beps^2-\bt^2\epsilon_2\right)\Lambda\right].
\eea
The $N=4$ superspace integration measure $dtd^4\theta$ is transformed
as
\be
\delta\, dtd^4\theta = 2i(\epsilon\cdot \bt + \bar\epsilon\cdot \theta)\,dtd^4\theta~.
\label{meastrans}
\ee
The covariant derivatives $D^i$, $\bar D_i$ transform as
\bea
\delta D^i = i\left[(2 +\alpha)(\epsilon\cdot\bt) +
\alpha (\theta\cdot \beps)\right]D^i
- 2i(1+\alpha)(\beps\cdot \bt)\bar D^i
 -2i\alpha \left[\epsilon^{(i}\bt_{k)} + \theta^{(i}\beps_{k)}\right] D^k,&&
\nonumber \\
\delta \bar D_i = i\left[(2 +\alpha)(\beps\cdot\theta) +
\alpha (\bt\cdot \epsilon)\right]\bar D_i
- 2i(1+\alpha)(\theta\cdot \epsilon)D_i
-2i\alpha \left[\epsilon_{(i}\bt_{k)} + \theta_{(i}\beps_{k)}\right]\bar D^k.&&
\label{barDtrans}
\eea
{}From these transformations it follows, in particular, that chiral
$N=4, d=1$ superfields
can be defined covariantly with respect to the superconformal transformations only
at $\alpha = -1$, i.e. in the case of the supergroup $SU(1,1\vert 2)$.

Since all other $D(2,1;\alpha)$ transformations appear in the anticommutator
of the conformal
and Poincar\'e supersymmetry generators, it is sufficient to require invariance
under these
two supersymmetries, when constructing invariant actions for the considered system.

For further use, we also give the explicit expressions for the variations
of our superspace
coordinates and superfields with respect to two $SU(2)$ subgroup. They are generated
by the left
action of the group element
\be
g_2=e^{iaV+i{\bar a}\bV} e^{ibT+i{\bar b}\bT}
\ee
and read
\bea\label{su2}
&&\delta\theta_1={\bar b}\bt^2-{\bar a}\theta_2,\quad \delta\theta_2
=-{\bar b}\bt^1+a\theta_1 ,\nonumber \\
&& \delta \Lambda=a+{\bar a}\Lambda^2,\; \delta\bLam={\bar a}+a\bLam^2,\quad
\delta\phi=i\left( a\bLam-{\bar a}\Lambda\right) .
\eea

\setcounter{equation}0
\section{N=4, d=1 ``hypermultiplet''}
The basic idea of our method is to impose the appropriate $D(2,1;\alpha)$ covariant
constraints on
the Cartan forms \p{cformdef}, \p{cforms}, so as to end up with some minimal
$N=4, d=1$ superfield
set carrying
an irreducible off-shell multiplet of $N=4, d=1$ supersymmetry.
Due to the covariance of the
constraints, the ultimate Goldstone superfields will support the corresponding
nonlinear realization of
the superconformal group  $D(2,1;\alpha)$.

Let us elaborate on this in some detail. It was the desire to keep $N=4, d=1$
Poincar\'e supersymmetry
unbroken that has led us to associate Grassmann coordinates
$\theta_i, \bar\theta^i$ with the
Poincar\'e supercharges in \p{coset} and fermionic Goldstone superfields
$\psi_i, \bar\psi^i$ with
the remaining four supercharges which generate conformal supersymmetry.
The minimal number of physical
fermions in an irreducible $N=4, d=1$ supermultiplet is four and it nicely matches
with the number
of fermionic Goldstone superfields in \p{coset} the first components of which
can so be naturally
identified with the fermionic fields of the ultimate Goldstone supermultiplet.
On the other hand, we can vary
the number of bosonic Goldstone superfields in \p{coset}: by putting some of them
equal to zero we can
enlarge the stability subgroup by the corresponding generators and so switch
to another coset with a
smaller set of parameters. Thus, for different choices of the stability
subalgebra the coset \p{coset}
will contain different numbers of the bosonic superfields, but always the
same number of fermionic superfields
$\psi_i,\bar\psi^i$. Yet, the corresponding sets of bosonic and fermionic
Goldstone superfields contain too many field components, and it is natural
to impose on
them the appropriate covariant constraints in order to reduce the number
of components as much as possible.
For preserving off-shell $N=4$ supersymmetry these constraints must be purely
kinematical, i.e.
not imply any dynamical restriction like equations of motion.

Some of the constraints just mentioned should express the Goldstone fermionic
superfields in terms
of spinor derivatives of the bosonic ones.On the other hand, as soon
as the first components of
the fermionic superfields $\psi_i,\bar\psi^k$
are required to be the only physical fermions, we are led to impose much stronger
condition
that {\it all} spinor derivatives of {\it all} bosonic superfields
are properly expressed in terms of
$\psi_i,\,\bar\psi^i\,$. Remarkably, these latter conditions
will prove to be
just the irreducibility constraints picking up irreducible $N=4$
supermultiplets.
In this and
next Sections we shall demonstrate how this procedure works for various cosets
which correspond
to placing some of the original coset bosonic generators
$D, V,\bV, V_3$ into
the stability subalgebra. Note that in the cases when the dilatation generator $D$
is among the coset generators (this is true for the options considered here
and in the next Section)
we should also keep the conformal boosts generator $K$ in the coset
on equal footing with
the translation generator $P$. This requirement is necessary
for the Cartan form $\omega_D$ in  \p{cforms}
to be separately covariant under the left action of $D(2,1;\alpha)$. Actually,
this is the standard way of doing
with nonlinear realizations of the $d=1$ conformal group
$SL(2,R) \sim SO(1,2)$ \cite{iklev1}.

We shall start with the most general case when the coset
\p{coset} contains all four bosonic
superfields $u,\varphi,\bar\varphi,\phi$. Looking at the structure of
the Cartan 1-forms
\p{cforms}, it is easy to find that the covariant constraints
which express all spinor covariant
derivatives of bosonic superfields in terms of the Goldstone fermions
amount to setting equal to zero
the spinor projections of these 1-forms
(these conditions are particular case of inverse Higgs effect \cite{IH}).
Thus, in the case at hand we impose the following constraints
\be\label{hyper1}
\left. \omega_D=\omega_V\left|=\bar\omega_V\right|=\omega_{V_3}\right|=0 \;,
\ee
where $|$ means restriction to spinor projections.
These constraints are manifestly covariant
under the whole supergroup $D(2,1;\alpha)$. They allow
one to express the Goldstone spinor superfields
as the spinor derivatives of the residual bosonic Goldstone superfields
$u, \Lambda, \bar\Lambda, \phi$ and  imply
some irreducibility constraints for the latter:
\bea\label{hyper2}
&& D^1 \Lambda = -2i\alpha \Lambda \left( \bpsi^1+\Lambda\bpsi^2\right),\;
D^1\bLam=-2i\alpha\left(\bpsi^2-\bLam\bpsi^1\right),\;
D^1 \phi =-2\alpha\left( \bpsi^1+\Lambda\bpsi^2\right),\nn
&& D^2\Lambda =2i\alpha\left(\bpsi^1+\Lambda\bpsi^2\right), \;
D^2\bLam=-2i\alpha\bLam\left( \bpsi^2-\bLam\bpsi^1\right),\;
D^2 \phi = 2\alpha \left(\bpsi^2-\bLam \bpsi^1\right), \nn
&& D^1 u =2i \bpsi^1 ,\;  D^2 u = 2i\bpsi^2, \quad \dot{u}=2z
\eea
(and c.c.). The irreducibility conditions in this and other cases which
we shall consider further arise due
to the property that the Goldstone fermionic superfields are simultaneously expressed
by \p{hyper2} in terms of spinor derivatives of different bosonic superfields. Then,
eliminating these spinor superfields, we end up with the relations between the spinor
derivatives of bosonic Goldstone superfields. To make these constraints most feasible,
it is advantageous to pass to the new variables
\be\label{hyper3}
q^1=\frac{e^{\frac{1}{2}(\alpha u -i\phi)}}{\sqrt{1+\Lambda\bLam}}\Lambda ,\quad
q^2=-\frac{e^{\frac{1}{2}(\alpha u -i\phi)}}{\sqrt{1+\Lambda\bLam}},\quad
{\bar q}_1=\frac{e^{\frac{1}{2}(\alpha u +i\phi)}}{\sqrt{1+\Lambda\bLam}}\bLam ,\;
{\bar q}_2=-\frac{e^{\frac{1}{2}(\alpha u +i\phi)}}{\sqrt{1+\Lambda\bLam}}\,.
\ee
In terms of these variables the irreducibility constraints acquire the manifestly
$SU(2)$ covariant form
\be\label{hyper4}
D^{(i}q^{j)}=0,\quad \bD{}^{(i}q^{j)}=0\,.
\ee
This $N=4,d=1$ multiplet was already considered, in the component
and $N=1$ superfield approaches,
in \cite{{CP},{GPS},{HP},{Hull1}} and also was recently studied in $N=4, d=1$ harmonic
superspace \cite{Olaf}.
Off shell it contains ${\bf 4}$ bosonic and ${\bf 4}$ fermionic fields and
{\bf no} auxiliary fields.
In this respect it resembles the $N=2, d=4$ hypermultiplet.
But in contrast to the $d=4$ case the constraints \p{hyper4} define an off-shell
multiplet in $d=1$. In fact, all other known $N=2, d=4$ supermultiplets
also have $N=4,d=1$
descendants. Their defining constraints follow from the $N=2, d=4$ ones by discarding
space-time spinor indices of the covariant derivatives. In the forthcoming Sections
we shall reproduce all these $N=4, d=1$ supermultiplets within our nonlinear
realizations framework.
Note that the $q^i$ supermultiplet can be considered as a fundamental
one since all its components
have the interpretation as Goldstone fields: ${\bf 4}$ fermions
are Goldstino for the nonlinearly
realized
conformal $N=4$ supersymmetry, while ${\bf 4}$ bosons are Goldstone fields
for the nonlinearly
realized
dilatations and $SU(2)$ transformations. All other irreducible multiplets
derived in next Sections
contain auxiliary fields which admit no immediate interpretation as Goldstone fields.

Let us note  that in the present approach it is easy to define the transformation
properties of $q^i$ under the group $D(2,1;\alpha)$ because we know the transformation
properties of $N=4, d=1$ superspace coordinates and all original Goldstone
superfields in
\p{coset}, while the constraints \p{hyper4} are covariant by construction.
In particular, the transformations of conformal supersymmetry read
\be\label{hyper5}
\delta q^i =2i\alpha\left( \bt^i\epsilon_j -\theta^i\beps_j\right) q^j \;.
\ee

Superconformally invariant superfield action of the sigma model type
can be easily found to be
\be\label{hyper6}
S^{(\alpha)}_q=\int dt d^4\theta \left( q\cdot {\bar q} \right)^{\frac{1}{\alpha}} \;.
\ee
The invariance of \p{hyper6} follows from the transformation properties of the
integration measure \p{meastrans} and $q\cdot{\bar q}$:
\be
\delta (q\cdot{\bar q})= -2i\alpha\left( \epsilon\cdot\bt +\beps\cdot\theta\right)
(q\cdot{\bar q})\;.
\ee
The case with $\alpha=-1$ should be considered independently, because
as a consequence of \p{hyper4}
we have
\be\label{hyper7}
D^iD_i \left( \frac{1}{q\cdot{\bar q}}\right)
= \bD_i \bD^i \left(\frac{1}{q\cdot{\bar q}}\right)=
[D^i,\bD_i] \left( \frac{1}{q\cdot{\bar q}}\right)=0
\ee
and therefore the action \p{hyper6} vanishes in this special case. Hence,
we instead consider the
action
\be\label{hyper8}
S^{(\alpha=-1)}_q= - \int dt d^4\theta\;
\frac{\mbox{ ln } \left( q\cdot{\bar q}\right)}{q\cdot{\bar q}} \;,
\ee
which is invariant up to a total derivative in the integrand. One can check
that $\frac{1}{1+\alpha}S^{(\alpha)}_q$ is
regular for any $\alpha$ and coincides with $S^{(\alpha=-1)}_q$ for $\alpha=-1$.
This situation is similar to the case of $N=4,d=1$ $V^{ij}$
supermultiplet considered in \cite{IKL} (see
next Section).

A more detailed discussion of possible actions for $q^i$ multiplet
can be found in \cite{Olaf}.
In particular, there exists a superpotential-type off-shell invariant which,
however, does not give
rise in components to any scalar potential for the physical bosons. Instead,
it produces a Wess-Zumino type term of first order in
time derivative. It can be interpreted as a coupling to a
four-dimensional background abelian
gauge field. The superpotential just mentioned admits a concise
manifestly supersymmetric superfield formulation
as an integral over an analytic subspace of $N=4, d=1$ harmonic superspace
\cite{Olaf}.

\setcounter{equation}0
\section{N=4, d=1 ``tensor'' multiplet}
This multiplet has been derived from a nonlinear realization of $D(2,1;\alpha)$ and
considered in detail
in \cite{IKL}. Here we shortly recall the basic points of
the construction of \cite{IKL}
as a particular
case of the general method described in the beginning of the previous Section.
The relevant formulas
will be needed for establishing relationships of this $N=4, d=1$ ``tensor''
multiplet with other ones.

The ``tensor'' multiplet corresponds to the choice $\phi = 0$
in the coset element \p{coset},
which amounts to transferring $U(1)\subset SU(2)$ into the stability subgroup.
Thus in this case
one deals with a supercoset involving $SL(2,R)\times \left[SU(2)/U(1)\right]$
as the bosonic manifold and, respectively,
with three bosonic Goldstone superfields $u, \Lambda, \bar\Lambda $.
In accord with our
general approach, we impose the following set of constraints:
\be\label{tensor1}
\omega_D=\omega_V\left|=\bar\omega_V\right|=0 \;.
\ee
Let us point out that now one cannot impose any constraints on the
Cartan form $\omega_{V_3}$
because it gets transforming inhomogeneously under $D(2,1;\alpha)$.
Explicitly, the set of constraints
\p{tensor1} reads
\bea\label{tensor2}
&& D^1 \Lambda = -2i\alpha \Lambda \left( \bpsi^1+\Lambda\bpsi^2\right),\;
D^1\bLam=-2i\alpha\left(\bpsi^2-\bLam\bpsi^1\right),\;
 D^1 u =2i \bpsi^1 ,\nn
&& D^2\Lambda =2i\alpha\left(\bpsi^1+\Lambda\bpsi^2\right), \;
D^2\bLam=-2i\alpha\bLam\left( \bpsi^2-\bLam\bpsi^1\right),\;
 D^2 u = 2i\bpsi^2 , \;\dot{u}=2z
\eea
(plus c.c.). After introducing a new $N=4$ isovector real superfield $V^{ij}$ ($V^{ij}=V^{ji}$ and
$\overline{V^{ik}}=\epsilon_{ii'}\epsilon_{kk'}V^{i'k'}$) via the identification
\bea\label{tensor3}
&& V^{11}=-i\sqrt{2}\,e^{\alpha u}\frac{\Lambda}{1+\Lambda\bLam} \; ,
\; V^{22}= i\sqrt{2}\,e^{\alpha u}\frac{\bLam}{1+\Lambda\bLam}\; , \;
V^{12}=  \frac{i}{\sqrt{2}}\,e^{\alpha u}\frac{1-\Lambda\bLam}{1+\Lambda\bLam} \;,
\nonumber \\
&& V^2 \equiv V^{ik}V_{ik} = e^{2\alpha u}~,
\eea
the irreducibility constraints for the bosonic superfields following
from \p{tensor2} can be cast
in the manifestly $SU(2)$-symmetric form
\be\label{tensor4}
D^{(i}V^{jk)} =0 \; , \quad \bD{}^{(i}V^{jk)} =0 \; .
\ee
The constraints \p{tensor4} could be obtained
by the direct dimensional reduction from the constraints
defining $N=2, d=4$ tensor multiplet \cite{tensord4} in which
one suppress the $SL(2,C)$ spinor indices of
$d=4$ spinor derivatives, thus keeping only the doublet indices of
the $R$-symmetry $SU(2)$ group.
This is the reason why we can call it
$N=4, d=1$ ``tensor'' multiplet.\footnote{An
alternative way to obtain $V^{ik}$ via a dimensional reduction
from $d=4$ is to start from
$N=1, d=4$ vector multiplet \cite{ismi}.
The $SL(2,C)$ group is reduced to $SU(2)$ and the indices
$\alpha, \dot\alpha$ becomes the doublet $SU(2)$ indices.
The superfield $V^{ik}$ then comes out
as the spatial component of the $d=4$ abelian gauge vector
connection superfield.} Of course,
in one dimension no any differential (notoph-type) constraints arise
on the components of the superfield $V^{ij}$. The constraints \p{tensor4}
leave in $V^{ik}$ the following independent superfield projections:
\be\label{tensor5}
 V^{ik}~, \; D^{i}V^{kl} =
-\frac{1}{3}(\epsilon^{ik}\chi^l + \epsilon^{il} \chi^k)~, \;
\bar D^{i}V^{kl} = \frac{1}{3}(\epsilon^{ik}\bar\chi^l
+ \epsilon^{il} \bar\chi^k)~, \;
D^i\bar D^k V_{ik}~,
\ee
where
\be
\chi^k \equiv D^iV_i^k~, \quad \bar\chi_k = \overline{\chi^k} = \bar D_i V^i_k~.
\ee
Thus its off-shell component field content is just $({\bf 3, 4, 1})$.
The $N=4, d=1$ superfield
$V^{ik}$ subjected to the conditions \p{tensor4}
was introduced in \cite{CR} and
later on rediscovered in \cite{ismi,bepa,stro2}.

Like in the case of the superfield $q^i$,
the $D(2,1;\alpha)$ superconformal transformations of
$V^{ij}$ can be deduced from the identification \p{tensor3}.
In particular, the conformal
supersymmetry acts as
\be
\delta V^{ij} = -2i\alpha\left[ (\epsilon\cdot\bt+\beps\cdot\theta)V^{ij} +
(\epsilon^{(i}\bt_k-\beps_k\theta^{(i})V^{j)k}+
(\epsilon_k\bt^{(i}-\beps^{(i}\theta_k)V^{j)k} \right].\label{tensor9}
\ee

Invariant superfield actions consist of a superfield kinetic term and
a superpotential \cite{IKL,ismi,Olaf}.
The superconformally invariant superfield kinetic term reads
\be\label{tensor12}
S_{V}^{(\alpha)}=\int dt d^4 \theta \left( V^2\right)^{\frac{1}{2\alpha}}.
\ee
As in the ``hypermultiplet'' case, for $\alpha = -1$ the action \p{tensor12}
vanishes and should be
replaced by
\be\label{tensor13}
S^{(\alpha=-1)}_V =
-\sfrac{1}{2}\int dt d^4 \theta \left( V^2\right)^{-\frac{1}{2}} \ln V^2.
\ee
The potential term can be written in a manifestly
$N=4$ supersymmetric form either with the help
of prepotential solving eqs. \p{tensor4} \cite{IKL},
or as an integral over the $N=4, d=1$
analytic harmonic superspace \cite{Olaf}.

As the last remark of this Section let us note that the explicit expressions
of the ``tensor'' multiplet \p{tensor3} and
the ``hypermultiplet'' \p{hyper3} in terms of initial Goldstone superfields give the hint
how to construct the former out of the latter. Indeed, from eqs. \p{tensor3}
and \p{hyper3}
it follows that
$V^{ij}$ can be represented as the following composite superfield
\be\label{tensor15}
\tV^{11}= -i\sqrt{2}\,q^1{\bar q}^1\; , \quad \tV^{22}
= -i\sqrt{2}\,q^2{\bar q}^2\;, \quad
\tV^{12}= -\frac{i}{\sqrt{2}}\left( q^1{\bar q}^2+ q^2{\bar q}^1 \right)\;.
\ee
One can check that, in consequence of the ``hypermultiplet'' constraints \p{hyper4},
the composite
superfield $\tV^{ij}$ automatically obeys \p{tensor4}. This is just the relations
established
in \cite{Olaf}.

The expressions \p{tensor15} supply a rather special solution
to the ``tensor'' multiplet constraints.
In particular, they express the auxiliary field of $\tV^{ij}$ through
the time derivative of
the physical components of $q^i$ which contains
no any auxiliary field.\footnote{This is a nonlinear version of
the phenomenon which is generic for $d=1$ supersymmetry and
was discovered at the linearized level in \cite{GR,pato}.}
As a consequence, the superpotential of $\tV^{ik}$ is a particular
case of the $q^i$ superpotential
which produces no genuine scalar potential for physical bosons and
gives rise for them only
to a Wess-Zumino type term
of the first order in the time derivative.

\setcounter{equation}0
\section{N=4, d=1 ``nonlinear'' multiplet}
In the previous two Sections we considered two cases
when the generator of dilatations $D$ is
placed into the coset along with some $SU(2)$ generators.
Alternative possibilities arise
if we place
the generator of dilatation $D$  into the stability subgroup
(together with the generator
of conformal boosts $K$). We firstly consider the case with
all generators of one $su(2)$ being present
in the coset. This case is singled out by setting $u=0, z=0$
in the general coset element \p{coset}.

The ``nonlinear'' multiplet we are going to construct
parametrizes the coset $SU(2)$ bosonic manifold. To express all spinor derivatives
from our set of bosonic superfields
$\varphi, \bar\varphi, \phi$ in terms of Goldstone fermions
$\psi^i, \bpsi^j$
we should impose the following set of constraints:
\be\label{nl1}
\omega_{V_3}\left|=\omega_V\left|=\bar\omega_V\right|\right.=0 \;.
\ee
They are covariant because the corresponding Cartan forms
are still homogeneously transformed
among themselves under the left action of $D(2,1;\alpha)$.
On the other hand, the Cartan form $\omega_D$
which enters the previous constraints \p{hyper1}, \p{tensor1}
now belongs to the stability subgroup algebra and
so transforms inhomogeneously. For this reason it cannot
be used for defining any covariant
constraint in the case under consideration.

Explicitly, the constraints \p{nl1} amount to the following set of equations
\bea\label{nl2}
&& D^1 \Lambda = -2i\alpha \Lambda \left( \bpsi^1+\Lambda\bpsi^2\right),\;
D^1\bLam=-2i\alpha\left(\bpsi^2-\bLam\bpsi^1\right),\;
 D^1 \phi =-2\alpha\left( \bpsi^1+\Lambda\bpsi^2\right) ,\nn
&& D^2\Lambda =2i\alpha\left(\bpsi^1+\Lambda\bpsi^2\right), \;
D^2\bLam=-2i\alpha\bLam\left( \bpsi^2-\bLam\bpsi^1\right),\;
 D^2 \phi = 2\alpha\left(\bpsi^2-\bLam \bpsi^1\right)
\eea
(and c.c.). They express 12 spinor derivatives of the bosonic superfield
$\varphi, \bar\varphi, \phi$
in terms of 4 fermions  $\psi^i, \bpsi^j$. Evidently,
this implies the existence
of additional
constraints on the bosonic Goldstone superfields. These constraints
can be put in a more concise form by passing to the new 4 by 4 matrix
variables $N^{ai}$:
\be\label{nl3}
 N^{11}=\frac{e^{-\frac{i}{2}\phi}}{\sqrt{1+\Lambda\bLam}}\Lambda,\;
N^{21}=\frac{e^{\frac{i}{2}\phi}}{\sqrt{1+\Lambda\bLam}},\;
N^{12}=-\frac{e^{-\frac{i}{2}\phi}}{\sqrt{1+\Lambda\bLam}},\;
N^{22}=\frac{e^{\frac{i}{2}\phi}}{\sqrt{1+\Lambda\bLam}}\bLam\;.
\ee
Here, the new doublet index $a$ is associated with some extra
global $SU(2)$ which commutes
with $D(2,1;\alpha)$ (and with the $N=4, d=1$ Poincar\'e superalgebra
$\subset D(2,1;\alpha)$).
The superfields $N^{ai}$ by construction obey the algebraic constraint
\be\label{nl4}
N^{ai}N_{ai}=2
\ee
which ensures the number of independent superfields to be three as it should be.
The additional irreducibility
constraints which follow from \p{nl2} can now be easily read off as
\be\label{nl5}
N^{a(i}D^j N^{k)}_a=0, \quad N^{a(i}\bD^j N^{k)}_a=0\; .
\ee
Comparing them with those of ref.\cite{nlin},
we recognize $N^{ai}$ as a $d=1$ analog of
the $N=2, d=4$ superfield which represents the nonlinear multiplet.

In order to reveal the component field content of $N^{ai}$,
it is convenient to pass to
another useful representation of the $d=1$ ``nonlinear'' multiplet
in terms of the following set of
superfields:
\be\label{nl5a}
X = e^{-i\phi} \Lambda , \; {\overline X}=e^{i\phi}\bLam ,\quad \phi \;.
\ee
The constraints \p{nl5} are rewritten as
\be\label{nl5b}
D^1 X=0, \quad \bD_2 X=0, \quad D^2\left(e^{i\phi}X\right)=-i D^1\phi,\quad
\bD_1\left(e^{i\phi}X\right) =i\bD_2\phi \;
\ee
(those including $\bar X$ follow by conjugation). Now it is clear that, due to \p{nl5b},
derivatives of each $N=4$ superfield with respect
to, say, $\theta_2,\bt^2$ can be expressed as derivatives
with respect to $\theta_1,\bt^1$ of
some other superfield. Therefore, only the $\theta_2=\bt^2=0$ components
of each superfield are
independent $N=2$ superfields. Moreover, the $N=2$ superfield $X|$
(with $\theta_2=\bt^2=0$)
is chiral. Thus, from $N=2$ point of view, the ``nonlinear'' multiplet contains one
general $N=2$ superfield $\phi$ and one chiral superfield $X$,
i.e. formally it has the same
off-shell field content $({\bf 3, 4, 1})$ as the ``tensor''
multiplet $V^{ij}$ of Section 4. However,
their
superconformal properties are radically different, and this is
the main distinction between
these two multiplets.
The transformation properties of $N^{ai}$ under conformal
supersymmetry \p{superconf} may
be directly determined using \p{nl3} and \p{superconf1}:
\be\label{nl6}
\delta N^{ai}= 2i\alpha \left( \epsilon^{(i}\bt^{j)} -
\beps^{(i}\theta^{j)}\right) N^a_j \;.
\ee
This has to be compared with the analogous transformation law
of $V^{ij}$, eq. \p{tensor9}.
Under the Poincar\'e supersymmetry the superfields $N^{ai}$ transform as scalars.
All other $D(2,1;\alpha)$
transformations appear in the anticommutator of the conformal
and the Poincar\'e supersymmetries.
Hence,
as in the previously considered cases,
it suffices to know explicit realization of these
two supersymmetries only. E.g., commuting \p{nl6}
with the Poincar\'e supersymmetry transformation, one
finds that the superfields $N^{ai}$ has the dilatation weight zero.
The same is true for physical
bosonic fields of this multiplet.

As regards the actions of the $d=1$ ``nonlinear'' multiplet,
the general off-shell action
\be\label{nl7}
S_N =\int dt d^4\theta L(N^{ai} ) \;,
\ee
where $ L(N^{ai} )$ is an arbitrary real function of $N^{ai}$,
is obviously invariant under $N=4$
Poincar\'e supersymmetry. Keeping in mind that $N^{ai}$
has zero dilatation weight, while the
integration measure in \p{nl7} has the weight $+1$,
one comes to the conclusion that superconformally invariant
action for $N^{ai}$ can be constructed only by coupling $N^{ai}$
to some dilaton-containing superfield.

The simplest example of $SU(2)\times SU(2)$ invariant action reads
\be\label{nl8}
S_1=-\int dt d^4\theta \mbox{ ln}(1+X\bX) =
-\int dt d^4\theta \mbox{ ln}(1+\Lambda\bLam)\;.
\ee
It is invariant under two commuting $SU(2)$ groups which one can define on
$N^{ai}$: the R-symmetry
$SU(2)$ acting on the indices $i$ and transforming superfields
$\Lambda$, $\bLam $ by the rule
\p{su2} and the second $SU(2)$ commuting with $N=4$ supersymmetry
(and with $D(2,1;\alpha)$)
and acting on the extra doublet indices $a$
\be
\delta_2 N^{ai} = \gamma^{(ab)}\,N_b^i\,.
\ee
In variables \p{nl5a} this second group acts as follows
\bea
&& \delta_2 X = \gamma^{11} - 2 \gamma^{12} X + \gamma^{22} (X)^2\,, \quad
\delta_2 \bar X = \gamma^{22} + 2 \gamma^{12} \bar X + \gamma^{11} (\bar X)^2\,, \nn
&&\delta_2 \phi = -2i\gamma^{12} + i \gamma^{22} X
-i \gamma^{11}\bar X\,.\label{Second}
\eea
Using the constraints \p{nl5b}, it is easy to check the invariance
of the action under both
these $SU(2)$ groups. Indeed,
\bea
\delta_1 \ln (1 + \Lambda\bLam) = a\bLam + \bar a \Lambda\,, \quad
\delta_2 \ln (1 +  X\bar X) = \gamma^{11}\bar X + \gamma^{22} X
\eea
and these variations vanish under the Berezin integral since, e.g., it follows
from \p{nl5b} that
$$
D^1 D^2 \Lambda = 0\,,
$$
while $X$ obeys the twisted chirality conditions.

If we define the physical components of the superfields $\phi, X, \bX$ as
\be\label{nlcomp}
 \phi|,\; X|,\; \bX|,\; \psi=D^1\phi|,\;\bpsi=-\bD_1 \phi|,\;\xi=D^1\bX|,\;
\bxi=-\bD_1 X|\;,
\ee
the action \p{nl8} takes the form (auxiliary fields vanish on shell):
\bea\label{nlcompaction}
S_1&=&\int dt\left[ {\dot\phi}^2-2\frac{i\dot\phi
({\dot X}\bX-X \dot\bX )-2{\dot X}{\dot\bX}}{1+X\bX}-
  \frac{X^2{\dot\bX}{}^2+2X\bX{\dot X}{\dot\bX}+
{\dot X}^2\bX{}^2}{(1+X\bX)^2}\right. \nn
&& \left. +2i\dot\psi\bpsi+2\frac{i\dot\xi \bxi
+X\xi{\dot\bpsi}+\bX{\dot\psi}\bxi}{1+X\bX}-
2i\frac{\xi\bxi( i{\dot\phi}+X{\dot\bX})}{(1+X\bX)^2} \right].
\eea
The bosonic part of \p{nlcompaction} can be concisely rewritten
in terms of $\phi, \Lambda,\bLam$:
\be
S_1^{bos}=\int dt\left[ \left(\dot\phi -
i\frac{\Lambda\dot\bLam-\dot\Lambda \bLam}{1+\Lambda\bLam}\right)^2 +
\frac{4\dot\Lambda\dot\bLam}{(1+\Lambda\bLam)^2} \right]. \label{Chir}
\ee
The term within the parentheses is just the $U(1)$ Cartan
form of the R-symmetry group
$SU(2)$ in the parametrization by $(\phi, \Lambda, \bLam)$ while the second term is
the $d=1$ pullback of the standard
metric on the coset $SU(2)/U(1)$. These two parts of the action \p{Chir} are
separately invariant under the left shifts of the R-symmetry $SU(2)$,
while their strict
ratio is fixed by the second $SU(2)$ invariance \p{Second} which rotates the $U(1)$
and $SU(2)/U(1)$ Cartan forms through each other.
Thus the action \p{Chir} actually defines a $d=1$
sigma model on the coset $SU(2)\times SU(2)/SU(2)_{diag}$, or, in other words,
a sigma model of principal chiral field on $SU(2)$. So the superfield action
\p{nl8} yields $N=4$ superextension of this sigma model.
In the general action \p{nl7} both these
$SU(2)$ symmetries can be broken. Note that the action \p{nl7}
and its particular case \p{nl8}
are trivially invariant under third $SU(2)$ group which
is contained in the original supergroup
$D(2,1;\alpha)$ and from the very beginning was placed into the
stability subgroup in our construction.
It acts only on the fermionic fields and rotates them through their conjugates.

Finally, we can write a composite version of the ``nonlinear'' multiplet
by expressing $N^{ai}$ in
terms of the ``angular'' part of the $d=1$ ``hypermultiplet''
\be\label{nlcomposite}
N^{ai}= \sqrt{2}\;\frac{q^{ai}}{\vert q \vert}\,,
\ee
where $q^{ai} \equiv (q^i, \bar q^i)$ and $\vert q \vert = \sqrt{q^{ai}q_{ai}}$.
As a consequence of \p{hyper4}, such a  composite  $N^{ai}$ automatically obeys the
constraints \p{nl5}. However, the representation \p{nlcomposite} is very restrictive,
since it expresses the auxiliary field of ``nonlinear'' multiplet
through the time derivative
of physical bosonic fields of the ``hypermultiplet''  $q^i$. It is worth noting
that this substitution
linearizes the sigma model action \p{Chir} when the auxiliary fields of $N^{ai}$
are retained. The
auxiliary fields term in the action becomes a kinetic term for the radial
part $\vert q \vert $ of $q^{ai}$ and it combines with the sigma model
Lagrangian for the angular
part of $q^{ai}$ in such a way  that one finally ends up with the
free $SU(2)\times SU(2)$ invariant
kinetic term for the $SO(4)$ vector $q^{ai}\vert $.

One can reformulate the ``nonlinear'' multiplet
in terms of analytic harmonic $N=4, d=1$ superfield
subjected to certain harmonic constraint \cite{Olaf},
in a full analogy with its $N=2, d=4$
prototype \cite{book}. Then, besides the sigma model action \p{nl7},
a superpotential term
can be set up for this multiplet
as an integral over the analytic harmonic $N=4, d=1$
superspace. We plan to perform a more detailed analysis of possible actions
for the $d=1$ nonlinear
multiplet and construction of the corresponding SQM models elsewhere.

\setcounter{equation}0
\section{N=4, d=1 ``nonlinear chiral'' multiplet}
In this Section we shall consider the last possibility for choosing
different bosonic
submanifolds in the coset \p{coset}. Namely, we shall keep in the coset
only bosonic superfields
parametrizing the sphere $S^2 \sim SU(2)/U(1)$.
Hence, we put  $u=0, z=0, \phi = 0$ in the coset element \p{coset}.
Now we have only two bosonic superfields $\varphi,\bar\varphi$
(or $\Lambda$, $\bar\Lambda$ related
to the former ones by the equivalence relation \p{Lambda}).
We impose the following  constraints:
\be\label{nlc1}
\omega_V\left|=\bar\omega_V\right|=0 \;.
\ee
Explicitly, after elimination of the fermionic Goldstone superfields,
these constraints
amount to the irreducibility conditions
\be\label{nlc2}
D^1\Lambda=-\Lambda D^2 \Lambda\,,    \quad \bD_2 \Lambda = \Lambda \bD_1 \Lambda
\quad (\mbox{and c.c.})\,.
\ee
The superfields $\Lambda, \bLam$ obeying  \p{nlc2} are recognized
as a nonlinear modification
of standard chirality constraints.
The crucial difference from the latter lies in that the
constraints \p{nlc2}
are covariant with respect to $D(2,1;\alpha)$ group for any $\alpha $,
while the chirality constraints
are covariant only for $SU(1,1|2) \sim D(2,1;-1)$.
The constraints \p{nlc2} leave
in $\Lambda, \bLam$ the following independent superfields projections:
\bea\label{nlc3}
&& \Lambda ,\; \bLam ,\; \psi=-D^1\bLam ,\;\bpsi=\bD_1 \Lambda , \; \nn
&& \xi=D^2 \Lambda ,\; \bar\xi = -\bD_2 \bLam , \; B=\bD_1 D^2 \Lambda , \;
{\bar B}=\bD_2 D^1 \bLam ,
\eea
which are just the irreducible ({\bf 2},{\bf 4},{\bf 2})
field content with $\Lambda , \bLam$ and $B, {\bar B}$
being the physical and auxiliary bosonic fields, respectively.

Like in the previous case of ``nonlinear'' multiplet,
the general sigma-model type action of $\Lambda, \bLam$
possesses only $N=4,d=1$ super Poincar\'e invariance and is given by
\be\label{nlc4}
S_\Lambda = \int dt d^4\theta\, L\left( \Lambda, \bLam\right) \;,
\ee
where $ L\left( \Lambda, \bLam\right)$ is an arbitrary real function
of $\Lambda, \bLam$.
The more restrictive case corresponds to preserving at least $U(1)$
symmetry generated by $V_3$.
The relevant action is
\be\label{nlc5}
S_2= \int dt d^4\theta\; f(\Lambda\bLam) \;,
\ee
where $f$ is a real function of the product $\Lambda\bLam$.
After passing to the components
\p{nlc3} and eliminating the auxiliary fields by their equations of motion,
action \p{nlc5} takes the
following form:
\bea\label{nlc6}
S_2 =&& \int dt \left\{
-4g \frac{\dot\Lambda\dot\bLam}{1+\Lambda\bLam} +
2ig \left( \psi\dot\bpsi + \xi\dot{\bar \xi}\right)+
2ig' \left(\dot\Lambda \bLam \psi\bpsi +\Lambda\dot\bLam \xi\bxi\right)-
2ig\frac{\xi\bpsi+\bxi\psi}{1+\Lambda\bLam}\right. \nn
&& +\left. \left[ \Lambda\bLam\left( 1+\Lambda\bLam\right)
\left( g'' -\frac{ (g')^2}{g}\right) +
\left( 1+\Lambda\bLam\right) g' +\frac{g}{1+\Lambda\bLam}\right]
\xi\bxi\psi\bpsi \right\}\;.
\eea
Here,
$$
g\equiv \left( 1+\Lambda\bLam\right)\left[ f'' \Lambda\bLam +f'\right]
$$
and prime means derivative with respect to $(\Lambda\bLam)$.

Let us remind that the superfields $\Lambda, \bLam$
do not contain dilaton (generator $D$ is now in
the stability subgroup). Therefore, like in the case of ``nonlinear'' multiplet,
it is impossible to construct
superconformally invariant actions of the sigma-model type out of $\Lambda, \bLam$ alone.
The superconformally
invariant action can likely be constructed only by coupling
the ``nonlinear chiral''  multiplet
to some other $N=4$ supermultiplet containing a dilaton among its field components.
On the other hand, it is easy to construct the action which
is invariant under the R-symmetry
$SU(2)$ transformations \p{su2}
\be\label{nlc7}
S_3= -\int dt d^4\theta \mbox{ ln}\left(1+ \Lambda\bLam\right)
\ee
(the general action \p{nlc4} and \p{nlc7} are trivially invariant under the group
$SU(2)$ which belongs to the stability subgroup and acts only on fermions).
In components the action \p{nlc7} takes the form
\be\label{nlc8}
S_2=\int dt\left[\frac{4\dot\Lambda\dot\bLam}{(1+\Lambda\bLam)^2}-
 2i\frac{\psi\dot\bpsi+\xi\dot\bxi}{1+\Lambda\bLam}+
2i\frac{\dot\Lambda\bLam\psi\bpsi+\Lambda\dot\bLam\xi\bxi+
\dot\bLam\xi\bpsi+\dot\Lambda\bxi\psi}{(1+\Lambda\bLam)^2} \right].
\ee
It is drastically simplified after passing to the new fermionic fields
\be\label{nlc9}
\widetilde\psi=\frac{\psi+\bLam \xi}{1+\Lambda\bLam}\;, \quad
\widetilde\xi=\frac{\xi-\Lambda \psi}{1+\Lambda\bLam}\;,
\ee
in terms of which it reads
\be\label{nlc10}
S_3=\int dt \left( \frac{4\dot\Lambda\dot\bLam}{(1+\Lambda\bLam)^2}-
 2i\widetilde\psi \dot{\overline{\widetilde\psi}}-
2i{\widetilde\xi} \dot{\overline{\widetilde\xi}} \right)\;.
\ee

We thus conclude that the action \p{nlc7} describes
a $N=4$ superextension of the $d=1$ $SU(2)/U(1)$
nonlinear sigma model.

The ``nonlinear chiral'' multiplet can be constructed as a composite one in terms
of the ``hypermultiplet'', ``tensor'' and even ``nonlinear'' $d=1$ multiplets.
The corresponding
expressions can be easily found
using \p{hyper3}, \p{tensor3} and \p{nl3}. In particular, in terms
of the ``hypermultiplet''
the superfields $\Lambda,\bLam$ can be expressed as:
\be\label{nlc11}
\Lambda = -\frac{q^1}{q^2},\quad \bLam=-\frac{{\bar q}_1}{{\bar q}_2}\;.
\ee
Of course, these realizations are very special since
the nonlinear chiral  multiplet contains
more auxiliary fields (just {\bf 2}) compared to
other multiplets (just {\bf 1} or {\bf 0}).
Therefore, in all
these realizations some of the auxiliary fields present
in $\Lambda,\bLam$ (or even all in the case
of ``hypermultiplet'') are expressed via the time derivatives of
physical components of $q^i, V^{ij}$  or $N^{ai}$.

Finally, we would like to point out that $N=2, d=4$ analog of
``nonlinear chiral'' multiplet is
unknown. A formal generalization of the constraints \p{nlc2}
to the $d=4$ case reads
\be\label{nlc12}
D^1_\alpha\Lambda=-\Lambda D^2_\alpha \Lambda ,
\quad \bD^1_{\dot\alpha} \Lambda = -\Lambda \bD^2_{\dot\alpha} \Lambda \;.
\ee
After passing to the new covariant derivatives defined as
\be\label{nlc13}
\cD^1_\alpha = D^1_\alpha + \Lambda D^2_\alpha, \quad {\overline \cD}^1_{\dot\alpha}=
 \bD^1_{\dot\alpha}+\Lambda \bD^2_{\dot\alpha}
\ee
the constraints \p{nlc12} take the very simple form
\be\label{nlc14}
\cD^1_\alpha \Lambda =0 ,\quad {\overline \cD}^1_{\dot\alpha} \Lambda=0 \;.
\ee
Formally, the derivatives \p{nlc13} look like covariant derivatives
in the harmonic $N=2, d=4$ superspace \cite{book},
with $\Lambda, \bLam$ being ``harmonics'' in a special gauge.
Then the constraints \p{nlc14} could be
treated as the Grassmann harmonic analyticity conditions
for $\Lambda, \bLam$. Hence, the
constraints \p{nlc12}
may be interpreted as the ones describing a special
nonlinear realization in harmonics superspace,
with harmonics becoming Goldstone $N=2$ superfields. Of course, the same
interpretation can be
done in the projective superspace \cite{rocek}. In this case the extra
complex variable, which is usually
introduced to define the set of anticommuting spinor derivatives
in the projective superspace,
should be
considered as an $N=2$ superfield obeying nonlinear chirality conditions.
The detailed discussion of such
a new type of $N=2, d=4$ supermultiplets is out of the scope of the present paper
and will be given elsewhere.

\setcounter{equation}0
\section{Special case: su(1,1$|$2) with central charges}
In the previous Sections we considered some  multiplets on which the supergroup
$D(2,1;\alpha)$ can be realized. They exist for arbitrary values of parameter
$\alpha$. The special
case $\alpha=-1$, when superalgebra $D(2,1;-1)$ is isomorphic to the
semi-direct sum of
$su(1,1|2)$ and $su(2)$, is also admissible.
The main peculiarity of this case is the
special form of superconformal invariant actions for the ``hypermultiplet'' and
``tensor'' multiplet.

Let us remind that the choice $\alpha=0$ is inadmissible with our definition
of the coset space \p{coset}
because the Cartan forms ${\hat\omega}_V, {\hat{\bar\omega}}_V,{\hat\omega}_{V_3}$
\p{cforms1} would be equal to zero in this case. Hence,
it would be impossible to relate the covariant derivatives of
scalar Goldstone superfields $\varphi,\bar\varphi, \phi$
to the fermions $\psi_i ,\bpsi^i$.
However, there is still a possibility to reach $\alpha=0$,
but for the properly contracted $D(2,1;\alpha)$.
Namely, let us redefine the $su(2)$ generators $V, \bar V, V_3$ as
\be\label{cc}
Z\equiv \alpha V,\quad \bZ\equiv \alpha \bV,\quad Z_3\equiv \alpha V_3 \;.
\ee
All the (anti)commutation relations of $D(2,1;\alpha)$ can be
rewritten in terms of $Z,\bZ,Z_3$.
Setting $\alpha=0$ in these relations leaves us with a $su(1,1|2)$ superalgebra
extended by three central charges. Indeed, from \p{alg1}-\p{alg4}
it immediately follows that for the choice
$\alpha=0$ the generators  $Z,\bZ,Z_3 $ commute
with everything (including themselves) but they still
appear in the cross-anticommutators of the Poincar\'e and
conformal supercharges:
\bea\label{alg4a}
&& \left\{ Q^i,S^j \right\} =-2\epsilon^{ij} \bT ,\;
\left\{Q^1 ,\bS_2 \right\} =2 \bZ , \; \left\{Q^1 ,\bS_1 \right\} =-2D-2 Z_3+2T_3 ,
               \nonumber \\
&& \left\{Q^2 ,\bS_1 \right\} =-2 Z, \; \left\{Q^2 ,\bS_2 \right\} =-2D +2 Z_3+2T_3 .
\eea
Now we can define a realization of this extension of $SU(1,1|2)$ by three central charges
in the following coset:
\be\label{ccoset}
g=e^{itP}e^{\theta_i Q^i+\bt^i \bQ_i}e^{\psi_i S^i+\bpsi^i \bS_i}
e^{izK}e^{iuD}e^{i\varphi Z+ i\bar\varphi \bZ}e^{\phi Z_3}
\ee
(as before, the other $su(2)$ generators $T, \bar T, T_3 $ are placed
into the stability subgroup).
Now the Cartan forms \p{cforms} are drastically simplified:
\bea\label{ccforms}
&& \omega_D = idu-2\left( \bpsi^i d\theta_i + \psi_i d\bt^i \right)
  -2iz d{\tilde t} \; , \nn
&& \omega_Z= id\varphi+2 \left[ \psi_2 d\bt^1 -\bpsi^1\left( d\theta_2-
       \psi_2 d{\tilde t}\right)\right], \;
 \bomega_{Z}= id\bar\varphi+2 \left[ \bpsi^2 d\theta_1 -\psi_1
    \left( d\bt^2-\bpsi^2 d{\tilde t}\right)\right],\nn
&& \omega_{Z_3}=d\phi+2\left[ \psi_1 d\bt^1 -\bpsi^1 d \theta_1-
   \psi_2 d\bt^2 +\bpsi^2 d \theta_2 +
 \left( \bpsi^1 \psi_1 -\bpsi^2 \psi_2\right)d{\tilde t}\right].
\eea
The superconformal transformations are generated by the left shifts of
the coset element \p{ccoset}
by the supergroup element \p{superconf}
$$
g_1= e^{\epsilon_i S^i +\beps^i \bS_i}~.
$$
Explicitly, they are
\bea\label{csuperconf1}
&& \delta t=-it \left( \epsilon \cdot\bt +\beps\cdot\theta \right) +
  \theta\cdot\bt \left( \epsilon\cdot\bt -\beps\cdot\theta \right),\;
\delta \theta_i= \epsilon_i t  + 2i
   \theta_i (\bt\cdot\epsilon) -i \epsilon_i (\theta\cdot\bt), \nn
&& \delta u= -2i \left( \epsilon \cdot\bt +\beps\cdot\theta\right),\;
\delta\phi=2\left[\beps^1\theta_1-\beps^2\theta_2 -
\epsilon_1\bt^1+\epsilon_2\bt^2\right], \nn
&&  \delta \varphi= 2i\left(\theta_2\beps^1-\bt^1\epsilon_2\right).
\eea

Now we shall list the $N=4, d=1$ supermultiplets which can be obtained by applying
our procedure to this central charge-extended $su(1,1\vert 2)$.
\vspace{0.4cm}

\noindent{\bf 7.1\ ``Hypermultiplet''}
\vspace{0.2cm}

As in the Section 3, we place all four bosonic
superfields $u,\varphi,\bar\varphi,\phi$ into the coset and
impose the following constraints
\be\label{suhyper1}
\left. \omega_D=\omega_Z\left|=\bar\omega_Z\right|=\omega_{Z_3}\right|=0 \;,
\ee
or, explicitly,
\bea\label{suhyper2}
&& D^1 \varphi = 0,\; D^1\bar\varphi=-2i\bpsi^2,\;
D^1 \phi =-2 \bpsi^1,\; D^1 u =2i \bpsi^1 , \nn
&& D^2\varphi =2i\bpsi^1, \;
D^2\bar\varphi=0,\;
D^2 \phi = 2\bpsi^2, \;  D^2 u = 2i\bpsi^2, \quad \dot{u}=2z
\eea
(and c.c.). After passing to the new variables
\be\label{suhyper3}
q^1=\varphi ,\quad
q^2=-\frac{1}{2}( u -i\phi),\quad
{\bar q}_1=\bar\varphi ,\;
{\bar q}_2=-\frac{1}{2}(u +i\phi),
\ee
the constraints acquire the familiar form
\be\label{suhyper4}
D^{(i}q^{j)}=0,\quad \bD{}^{(i}q^{j)}=0.
\ee
Surely, no $SU(2)$  group realized on the doublet
indices of $q^i$ and $D^i$ is present now,
despite the fact that the constraints look $SU(2)$-covariant.
The simplest $SU(1,1|2)$ invariant action can be written as
\be\label{suhyperaction}
S=\int dt d^4\theta  \; e^{-(q^2+{\bar q}_2)}(q^2+{\bar q}_2) \equiv
- \int dt d^4\theta  \; e^u u  \;.
\ee
\vspace{0.4cm}

\noindent{\bf 7.2\ ``Tensor'' multiplet}
\vspace{0.2cm}

In this case the constraints formally coincide with  \p{tensor1}, but look much
simpler when rewritten in the explicit form:
\bea\label{sutensor}
&& D^1 \varphi = 0,\; D^1\bar\varphi=-2i\bpsi^2,\;
 D^1 u =2i \bpsi^1 , \nn
&& D^2\varphi =2i\bpsi^1, \;
D^2\bar\varphi=0,\;
 D^2 u = 2i\bpsi^2
\eea
(and c.c.). Once again, we can introduce a new $N=4$  superfield $V^{ij}$ such that
$V^{ij}=V^{ji}$ and $\overline{V^{ik}}=\epsilon_{ii'}\epsilon_{kk'}V^{i'k'}$
via the identification
\be\label{sutensor3}
V^{11}=-i\sqrt{2}\varphi \; ,
\; V^{22}= i\sqrt{2}\bar\varphi\; , \;
V^{12}=  \frac{i}{\sqrt{2}}u~,
\ee
and rewrite the constraints \p{sutensor} as
\be\label{sutensor4}
D^{(i}V^{jk)} =0 \; , \quad \bD{}^{(i}V^{jk)} =0 \; .
\ee
Thus, we have the same ``tensor'' multiplet as before but
with much simpler expression
for it in terms of coset fields. The price we have
to pay for this simplicity is the lost
of $su(2)$ invariance.

The invariant action for this case in terms of $N=2$ projections of $N=4$ superfields
\be\label{ppp}
{\tilde u}= u| ,\quad \lambda=\varphi|, \quad \bar\lambda=\bar\varphi| \;,
\ee
where $|$ means restriction to $\theta^2=\bt_2=0$, has the following form
\be\label{sutensor5}
S=
-\sfrac{1}{2}\int dt d^2 \theta\left[  e^{\tilde u}
\left(\bD D {\tilde u} +D\blam \bD\lambda\right)  +
  \kappa \mbox{ ln}\left( \frac{{\tilde u}+
\sqrt{{\tilde u}{}^2+4\lambda\blam}}{2}\right)  \right] .
\ee
It contains the kinetic and potential terms in analogy with the $SU(2)$
``tensor'' multiplet of
the Section 4.
\vspace{0.4cm}

\noindent{\bf 7.3\ ``Nonlinear'' and ``nonlinear chiral'' multiplets}
\vspace{0.2cm}

The Cartan forms \p{ccforms} for the case we are considering
do not contain any nonlinearities.
Hence, we could expect that ``nonlinear'' multiplets will
actually become linear. This is indeed so.

For these ``nonlinear'' multiplets we obtain the following constraints
\bea\label{sunl}
&& D^1 \varphi = 0,\; D^1\bar\varphi=-2i\bpsi^2,\;
D^1 \phi =-2 \bpsi^1, \nn
&& D^2\varphi =2i\bpsi^1, \;
D^2\bar\varphi=0,\;
D^2 \phi = 2\bpsi^2
\eea
(and c.c.). They can be rewritten just as \p{sutensor4}.
Therefore, ``nonlinear'' multiplet for the
contracted superalgebra coincides with the ``tensor'' one.
But their transformation properties
are still radically different, because the ``tensor'' multiplet contains
the dilaton $u$ while
in the ``nonlinear'' case we deal only with Goldstone fields
for the central charge
generators.

A similar linearization takes place for the nonlinear chiral multiplet.
Indeed, the constraints now read
\be\label{sunlc}
D^1 \varphi = 0,\; \bD_2\varphi =0
\ee
and define a sort of twisted chiral $d=1$ multiplet.

Any Lagrangian function of these superfields will respect manifest off-shell $N=4, d=1$
supersymmetry.
In these cases, as distinct from the options considered in the previous Sections,
one cannot
construct out of these superfields alone not only superconformally invariant, but
also $SU(2)$ invariant actions
(though still persists the ``trivial'' $SU(2)$ realized on fermions only).

\vspace{0.4cm}

\noindent{\bf 7.4\ ``Old'' multiplets}
\vspace{0.2cm}

Here we mention two additional possibilities which exist for the central charge
extended $su(1,1|2)$ superalgebra. Actually, one of them exists as well in the general
$D(2,1;\alpha)$ case.

First of all, we could consider the case when the set of bosonic Goldstone superfields
in the coset \p{ccoset} includes only dilaton $u$ and one extra Goldstone superfield
$\phi$ associated with the central charge $Z_3$.
The set of constraints in this case
\be\label{suchiral}
D^1 \phi =-2 \bpsi^1,\; D^1 u =2i \bpsi^1 , \quad
D^2 \phi = 2\bpsi^2, \;  D^2 u = 2i\bpsi^2  \quad (\mbox{and c.c.})
\ee
defines the ordinary $N=4$ chiral superfield $u + i\phi\,$.
Let us remind that the chirality
conditions are compatible only with the $su(1,1|2)$ superalgebra.
In more detail such a supermultiplet
and the corresponding actions have been considered in \cite{ikl2}.

The last possibility corresponds to retaining the single dilaton $u$
in the bosonic part of the
coset \p{ccoset}.
In this case no any constraints appear since four Goldstone fermions
are expressed through four
spinor derivatives of $u$. As was shown in \cite{ikl2}, one should impose
some additional irreducibility constraints on dilaton $u$
\be\label{last}
D^i D_i\; e^{-\alpha u}= \bD{}_i\bD^i\; e^{-\alpha u}=
\left[ D^i,\bD_i \right] e^{-\alpha u} =0
\ee
in order to pick up in $u$ the minimal off-shell field
content ({\bf 1},{\bf 4},{\bf 3}).
Once again, the detailed discussion of this case
can be found in \cite{ikl2}.

In fact, the same multiplet can be derived from the nonlinear
realization of $D(2,1;\alpha)$ in its
coset \p{coset}, with all bosonic superfields except the dilaton being equal to zero.
This clearly corresponds to placing both internal $SU(2)$ groups
into the stability subgroup. The
only constraint is
\be
\omega_D\vert = 0\,,
\ee
and it serves to trade the Goldstone fermions for the spinor derivatives
of $u$ without
implying any additional constraints for $u$. One can further constrain
$u$ by \p{last} and check
the covariance of this set of constraints under the whole $D(2,1;\alpha)$.
This example shows
that our method is directly applicable for deriving only those irreducible $N=4, d=1$
superfields which are defined by constraints of the first order in spinor derivatives.
On the other hand, we could re-derive the multiplet $u$ from our $V^{ik}$ discussed in
\cite{IKL} and Section 4. Indeed, one can construct the composite superfield
\be
e^{-\alpha\tilde{u}} = \frac{1}{\sqrt{V^2}}\,, \label{last1}
\ee
which obeys just the constraints \p{last} as a consequence of \p{tensor4}.
The relation \p{last1} is
of the same type as the previously explored substitutions
\p{tensor15} or \p{nlcomposite} and
expresses two out
of the three auxiliary fields of $\tilde u$
via physical bosonic fields of $V^{ik}$ and time derivatives
thereof. The $D(2,1;\alpha)$ invariant action
for the superfield $u$ at $\alpha \neq -1$ and $\alpha = -1$
can be then obtained by the substitution $\sqrt{V^2}
\rightarrow e^{\alpha u}$ in \p{tensor12}
and \p{tensor13}
\be
S_u^{(\alpha)} = \int dt d^4\theta\, e^{u}\,, \quad S_u^{(\alpha = -1)}=
\int dt d^4\theta \, e^{u}\, {u}\,.
\ee
Note that in \cite{ikl2} only the action for the $SU(1,1\vert 2)$ case
(i.e. with $\alpha = -1$) was considered.

\setcounter{equation}0
\section{Summary and concluding remarks}
In this paper we showed that a lot of off-shell $N{=}4, d{=}1$ supermultiplets with
irreducibility constraints of first order in spinor derivatives
can be self-consistently
derived from nonlinear realizations of the most general $N{=}4, d{=}1$ superconformal
group $D(2,1;\alpha)$ in its appropriate coset superspaces. Multiplets with
irreducibility constraints of higher order
in spinor derivatives can be derived from
these basic ones via a chain of proper substitutions.
In this way we derived most of the previously known multiplets
as well as two new multiplets which have not yet been exploited in
constructing $N{=}4$ supersymmetric SQM models.

Our results are summarized in the Table below.
\vspace{0.5cm}

\begin{tabular}{|l|c|c|c|c|c|}
\hline
multiplet & content & R symmetry coset & dilaton & $\alpha$ & superfield \\
\hline
``old tensor'' & ({\bf 1,4,3}) &  -- & yes & any & $u$  \\
  chiral & ({\bf 2,4,2}) &  central charge & yes & $0, -1$ & $\phi, \bar\phi$  \\
nonlinear chiral& ({\bf 2,4,2}) &  $su(2)/u(1)$ & no & any & $\Lambda,
\bar\Lambda$  \\
tensor & ({\bf 3,4,1}) & $su(2)/u(1)$ & yes & any & $V^{ij}$  \\
nonlinear & ({\bf 3,4,1}) & $su(2)$ & no & any & $N^{ia}$  \\
hypermultiplet & ({\bf 4,4,0}) &  $su(2)$ & yes & any & $q^{ia}$\\
\hline
\end{tabular}
\vspace{0.5cm}

Using the same method, we also derived $N{=}4$
multiplets associated with nonlinear realizations of the
supergroup $SU(1,1\vert 2)$ modified
by three central charges. Our method automatically yields
the superconformal transformation properties
of the final irreducible superfields, which is helpful
in constructing their superconformally
invariant actions. However, these superfields can be used
equally well for constructing
actions which in general
exhibit only Poincar\'e supersymmetry.

As an interesting project for further study we mention
the explicit construction
of SQM models associated with the new $N{=}4, d{=}1$ supermultiplets found here.
Another
promising task includes generalizing these techniques
to $N{>}4, d{=}1$ supersymmetries, classifying
with its help the corresponding supermultiplets
and constructing examples of the SQM models
associated with these supermultiplets. At present
not too much is known about SQM models
with such a large amount of supersymmetry. As a first step
towards this goal one may attack the
case of $N{=}8, d{=}1$ supersymmetry \cite{underway}.

Finally, let us briefly argue why there is no need to also consider the situation
where generators from both $SU(2)$ groups $\left(V, {\bar V}, V_3\right)$ and
$\left( T,{\bar T}, T_3\right)$ are present in the coset \p{coset}.
For generic $\alpha$ there are not too many possibilities
to place generators from both $SU(2)$ groups into the coset.
Indeed, we have only four fermionic superfields in the
game. Therefore, the maximal number of bosonic superfields must be
less then or equal to four.
This leaves only three possibilities for the bosonic coset:\\
a)  $SU(2)\times SU(2)/SU(2)_{diag}$;\\
b)  $SU(2)\times SU(2)/SU(2)_{diag}$ plus dilaton $u$;\\
c)  $\left[SU(2)/U(1)\right] \times \left[SU(2)/U(1)\right].$\\
Let us dwell on the first case. At the linearized level
the corresponding constraints read
\bea
&& i D^1 \phi=0\,,\, iD^2 \phi+\alpha \bpsi^1=0\,, \,
i\bD_1\phi-\alpha \psi_2+(1+\alpha)\bpsi_1=0\,,\,
i\bD_2 \phi+(1+\alpha)\bpsi_2=0\,,\nn
&&D^1\phi_3+(1+2\alpha)\bpsi^1=0,\;D^2\phi_3+\bpsi^2=0,\;
\bD_1\phi_3-\psi_1=0,\;\bD_2\phi_3-\psi_2=0, \nonumber
\eea
where the $N{=}4$ superfields $\phi,\bphi,\phi_3$ parametrize the coset
$SU(2)\times SU(2)/SU(2)_{diag}$. It is easy to observe that
these constraints together with their complex conjugates imply
$$ \dot\phi=\dot{\bphi}=\dot{\phi}_3=0 \;.$$
Apparently, these constraints are too strong, and the corresponding case
should be excluded from our
consideration. The same is true for the other two possibilities mentioned above.

We close by noting that one can construct
``mirror'' $N{=}4, d{=}1$ multiplets
by switching between the two internal $SU(2)$
groups: place the one acting on indices $i,k$ into the stability subgroup
while lifting the other one up into the coset.
Clearly, the resulting superfield constraints will look the same as before,
modulo a different splitting of the $N{=}4, d{=}1$ superspace coordinates
and spinor derivatives into the $SU(2)$ doublets. New interesting
possibilities may arise from actions which include both
types of multiplets simultaneously.

\section*{Acknowledgments}
This work was partially supported by INTAS grant No 00-00254,
RFBR-DFG grant No 02-02-04002, grant DFG No 436 RUS 113/669, RFBR grant
No 03-02-17440 and a grant of the Heisenberg-Landau programme.
E.I. and S.K. thank the Institute
for Theoretical Physics of the University of Hannover
for the warm hospitality extended to
them during the course of this work.

\end{document}